%% file: main.tex
\documentclass[conference,10pt]{IEEEtran}

\usepackage{algpseudocode}
\algtext*{EndWhile}
\algtext*{EndIf}
\algtext*{EndFor}
\algtext*{EndProcedure}
\usepackage{amsmath}
\usepackage{amssymb}
\usepackage{array}
\usepackage{algorithm}
\usepackage{bm}
\usepackage[bookmarks=false, hidelinks]{hyperref}
\usepackage[noadjust]{cite}
\usepackage[font=small]{caption}
\usepackage{color}
\usepackage{comment}
\usepackage{dblfloatfix}
\usepackage[inline]{enumitem}
\usepackage{epsfig}
\usepackage{etoolbox}
\usepackage{fancybox}
\usepackage{float}
\usepackage{fixltx2e}
\usepackage{graphicx}
\usepackage{listings}
\usepackage{mathrsfs}
\usepackage{multirow}
\usepackage{multicol}
\usepackage{pifont}
\usepackage{placeins}
\usepackage{rotating}
\usepackage{setspace}
\usepackage{soul}
\usepackage[hang, tight]{subfigure}
\usepackage{threeparttable}
\usepackage{times}
\usepackage{url}
\usepackage{upgreek}
\usepackage{verbatim}
\usepackage{wrapfig}
\usepackage[usenames,dvipsnames]{xcolor}
\usepackage{authblk}
\usepackage[]{siunitx}
\usepackage[super]{nth}
\usepackage{tabularx}
\usepackage{supertabular}
\usepackage{tikz}
\usepackage{bbm}
\usepackage{blindtext}
\usepackage{makecell}
\usepackage{adjustbox}
\usepackage{afterpage}
\usepackage{booktabs}

\graphicspath{{./_fig/}}

\makeatletter
\renewcommand\footnoterule{%
  \kern-3\p@
  \hrule\@width0.4\columnwidth
  \kern2.6\p@}
  \makeatother

\usepackage{graphicx}
\usepackage{adjustbox}
\def\halfcheckmark{\tikz\draw[scale=0.4,fill=black](0,.35) -- (.25,0) -- (1,.7) -- (.25,.15) -- cycle (0.75,0.2) -- (0.77,0.2)  -- (0.6,0.7) -- cycle;}

\def\fullcheckmark{\tikz\draw[scale=0.4,fill=black](0,.35) -- (.25,0) -- (1,.7) -- (.25,.15);}

\newcommand*\circled[1]{\tikz[baseline=(char.base)]{
            \node[shape=circle,fill,inner sep=1pt,scale=0.8] (char) {\textcolor{white}{#1}};}}

\usepackage{bbding}

\usepackage{float}

\usepackage{color}

\definecolor{gg}{RGB}{0, 129, 35}
\definecolor{orangeShallow}{RGB}{255,190,0}

\usepackage[hang,flushmargin]{footmisc}

\begin{document}


\title{
Spec2SVA-Eval: Evaluating LLMs for Assertion Generation from Design Specification
}

\title{
Spec2Assert: Generating and Evaluating Assertion from Natual Language Specification
}

\title{Spec2Assert: Benchmarking LLM-based Assertion Generation from Natural Language Specification}

\title{AssertLLM: Hardware Verification Assertion Generation and Evaluation from Design Specification via Multi-LLMs}


\title{{\huge{AssertLLM: Hardware Verification Assertion Generation and Evaluation from Design Specification via Multi-LLMs}}}

\title{{\huge{AssertLLM: Generating and Evaluating Hardware Verification Assertions from Design Specifications via Multi-LLMs}}}

\author[]{ \fontsize{11}{11}\selectfont Wenji Fang$^{1,2}$, Mengming Li$^1$, Min Li, Zhiyuan Yan$^2$,  Shang Liu$^1$, Hongce Zhang$^{1,2}$\textsuperscript{*}, Zhiyao Xie$^1$\textsuperscript{*}\vspace{-5pt}}

\affil[]{\fontsize{10}{10}\selectfont $^1$Hong Kong University of Science and Technology, \\$^2$Hong Kong University of Science and Technology (Guangzhou)\vspace{-6pt}}

\affil[]{$\textsuperscript{*}$Corresponding Author: \{hongcezh, eezhiyao\}@ust.hk\vspace{-12pt}}

\maketitle
\thispagestyle{plain}
\pagestyle{plain}

\input{_txt/abstract}

\input{_txt/1_introduction}
\input{_txt/2_preliminaries}

\input{_txt/3_methodology}

\input{_txt/4_experiment}

\input{_txt/5_discussion}

\input{_txt/6_conclusion}

\newpage
\bibliographystyle{IEEEtran}
\bibliography{spec, references_1, references_2, assertion}
\end{document}

%% file: _txt/abstract.tex
\begin{abstract}
Assertion-based verification (ABV) is a critical method for ensuring design circuits comply with their architectural specifications, which are typically described in natural language. 
This process often requires human interpretation by verification engineers to convert these specifications into functional verification assertions.
Existing methods for generating assertions from natural language specifications are limited to sentences extracted by engineers, discouraging its practical application.
In this work, we present AssertLLM, an automatic assertion generation framework that processes complete specification files. AssertLLM breaks down the complex task into three phases, incorporating three customized Large Language Models (LLMs) for extracting structural specifications, mapping signal definitions, and generating assertions. 
Our evaluation of AssertLLM on a full design, encompassing 23 I/O signals, demonstrates that 89\% of the generated assertions are both syntactically and functionally accurate.\looseness=-1

\end{abstract}

%% file: _txt/1_introduction.tex
\section{Introduction}\label{sec:intro}

Hardware functional verification is critical in the VLSI design flow, primarily addressing whether an implementation adheres to its specification. For instance, the register-transfer level (RTL) design of a processor must comply with the given instruction set architecture (ISA) specification. Typically, the architects first develop the specifications in a natural language document. Subsequently, RTL designers translate these specifications into RTL code, while the verification engineers are responsible for checking the functional correctness of the RTL designs according to the specifications.

During the verification process, assertion-based verification (ABV)~\cite{witharana2022survey} is a widely adopted technique, which utilizes assertions crafted from specifications to verify the functional behavior of RTL designs. ABV can be conducted either through simulation with testbenches or using formal property verification (FPV) techniques. Temporal logic, particularly SystemVerilog Assertions (SVA), is commonly employed for specifying these properties. However, a significant challenge in ABV is the generation of sufficient, high-quality assertions. Currently, designing SVAs manually is a time-consuming and error-prone task, demanding unignorable human effort.

To address this challenge, research has focused on generating SVAs automatically. The automatic approaches can be categorized into two types: dynamic mining from simulation traces and static analysis of specifications. 
Dynamic methods~\cite{germiniani2022harm, danese2017team, vasudevan2010goldmine} generate assertions by combining simulating test traces and static analysis of design constraints. However, a critical limitation of dynamic methods is that both the generation and evaluation of assertions are on the same RTL design without referring to a golden reference model. This could lead to the generation of incorrect SVAs due to flaws in the RTL design, which these methods might not detect.
On the other hand, existing static methods depend either on the pre-defined templates~\cite{orenes2021autosva, fang2023r} or on machine learning (ML) technologies~\cite{kande2023llm, orenes2023using, sun2023towards, harris2016glast,krishnamurthy2019controlled,zhao2019automatic,krishnamurthy2019ease,frederiksen2020automated,keszocze2019chatbot,parthasarathy2021spectosva,aditi2022hybrid}.
The template-based methods also require a deep understanding of the design function to fill in the templates. As for the ML-based methods, they explore both traditional natural language processing (NLP) and emerging Generative AI techniques like Large Language Models (LLMs). We further categorize the existing static ML-based methods based on their application in different design phases: the RTL and pre-RTL stages.

Table~\ref{tbl:related_work} details these ML-based SVA generation methods in both the RTL stage and the pre-RTL stage. During the RTL stage, the process typically involves using LLMs to process both human-written specification sentences and the RTL design to generate SVAs describing security or functional properties~\cite{kande2023llm, orenes2023using, sun2023towards}. However, similar to the dynamic methods, inaccuracies in RTL implementations could result in flawed SVAs.

\begin{figure}[!t]
  \centering
  \includegraphics[width=1\linewidth]{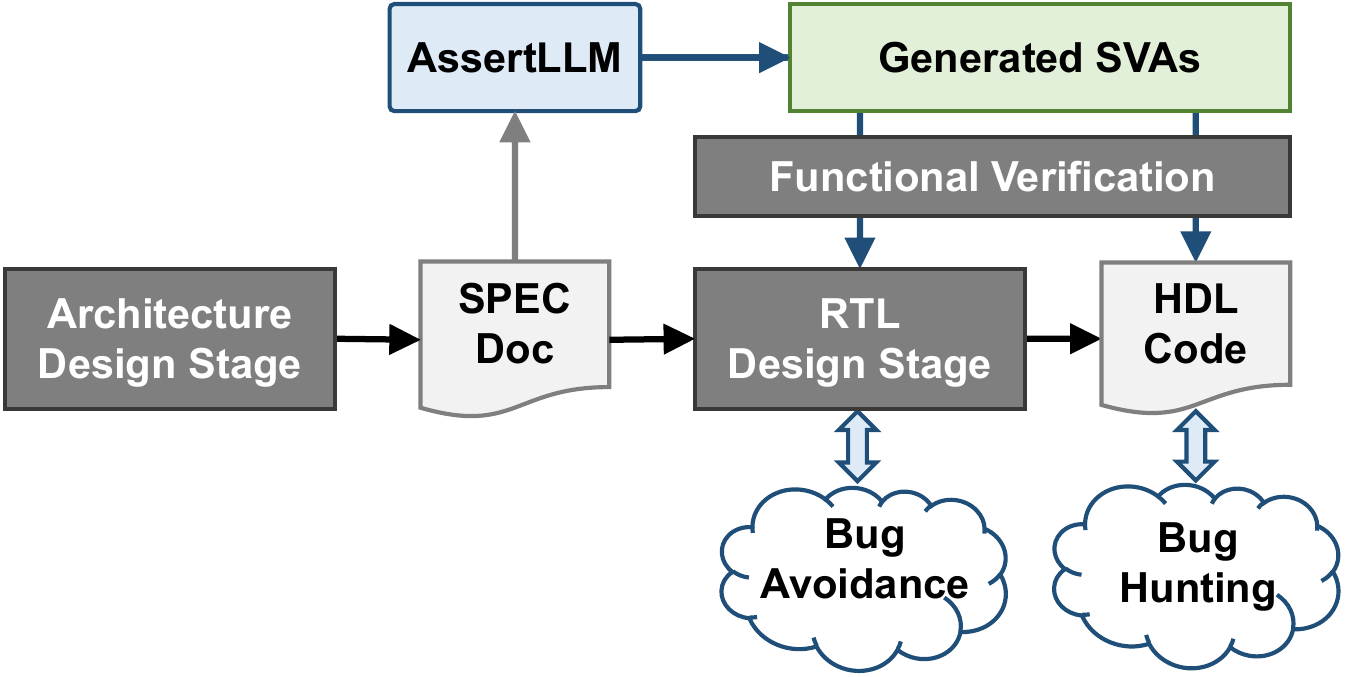}
  \caption{AssertLLM in VLSI design and verification flow. AssertLLM automatically generates SVAs from natural language specifications, facilitating functional verification for both bug avoidance and bug hunting.}
  \label{fig:motivation}
  \vspace{-.2in}
\end{figure}

\begin{table*}[!t]
\resizebox{1\textwidth}{!}{
\begin{tabular}{c||c|c|c|c|c|c|c}

\toprule
\multirow{2}{*}{\textbf{Stage}} & \multirow{2}{*}{\textbf{Works}}             & \multirow{2}{*}{\textbf{\begin{tabular}[c]{@{}c@{}}Generation \\      Method\end{tabular}}} & \multicolumn{2}{c|}{\textbf{NL Specification}}                                                                                          & \multicolumn{3}{c}{\textbf{Evaluation}}                                                                                             \\ \cline{4-8}
                                &                                             &                                                                                             & \textbf{Auto Extract.}                                                                          & \textbf{Source}     & \textbf{Full Design}           & \textbf{Target}                                                                                             
                                & \begin{tabular}[c]{@{}c@{}}\textbf{Open-Source} \\ \textbf{Benchmark}\end{tabular}
                                \\ \hline \hline
\multirow{2}{*}{\textbf{RTL}}            & \cite{kande2023llm}                                     & \multirow{2}{*}{LLM-based}                                                                  & \multirow{2}{*}{\XSolidBrush}   & \multirow{2}{*}{\begin{tabular}[c]{@{}c@{}}\textcolor{red}{Sentences} \\      from Engineers\end{tabular}   }

& \multirow{2}{*}{\XSolidBrush} & Security                                                                                                     & \XSolidBrush                    \\ \cline{2-2} \cline{7-8}
                                & \cite{orenes2023using, sun2023towards}                                        &                                                                          &     &                                                                                               &                             & Function                                                      & \halfcheckmark$^\star$         \\ \hline
\multirow{2}{*}{\textbf{Pre-RTL}}        & \cite{harris2016glast,krishnamurthy2019controlled,zhao2019automatic,krishnamurthy2019ease,frederiksen2020automated,keszocze2019chatbot,parthasarathy2021spectosva,aditi2022hybrid} & NLP-based                                                                & \XSolidBrush                   & \begin{tabular}[c]{@{}c@{}}\textcolor{red}{Sentences} \\      from SPEC file\end{tabular}                 &\fullcheckmark           &  \begin{tabular}[c]{@{}c@{}}Function \textcolor{red}{(specialized}\\       \textcolor{red}{checkers/ artificial cases)}\end{tabular}   & \XSolidBrush                    \\ \cline{2-8} 
                                & Ours                                        & LLM-based                                 & \fullcheckmark                                                  &\textcolor{gg}{Entire} SPEC file       &\fullcheckmark           & \begin{tabular}[c]{@{}c@{}}Function    \\     \textcolor{gg}{ (general benchmark)}\end{tabular}  & \fullcheckmark   \\ \bottomrule                
\end{tabular}

}
\begin{tablenotes}\footnotesize
\item $^\star$ Work~\cite{orenes2023using} only open-source the result on a FIFO, while work~\cite{sun2023towards} only demonstrate based on tiny designs such as FSM and DFF.
\end{tablenotes} 
\caption{Existing works on generating SVAs from natural language specifications. AssertLLM is the first work that can handle full-size specification files and generate comprehensive types of SVAs for each architectural signal. We also propose the first open-source benchmark for assertion generation and evaluation from natural language specifications.}
\label{tbl:related_work}
\vspace{-.1in}
\end{table*}

When it comes to the pre-RTL stage, with the natural language specification document finalized, RTL designers proceed to implement behavior satisfying this golden specification.
Numerous studies~\cite{harris2016glast,krishnamurthy2019controlled,zhao2019automatic,krishnamurthy2019ease,frederiksen2020automated,keszocze2019chatbot,parthasarathy2021spectosva,aditi2022hybrid} have employed NLP techniques to generate SVAs from sentences extracted by humans. These works focused on processing sentences identified from a comprehensive document of specification.
However, specification extraction requires tremendous human efforts, and the NLP-based generation process faces challenges in generalizing across diverse grammatical variations. 
Additionally, the evaluation of SVAs generated through these methods typically depends on design-specific checkers, such as protocol and processor checkers, and is therefore difficult to extend to other design types.

Here we summarize three key challenges that currently hinder the practical application of SVA generation from natural language specifications:
\begin{enumerate}
    \item Natural language VLSI specifications are inherently unstructured and are hard to be directly used for assertion generation.
    \item Even with structured specifications, translating natural language into assertions remains a highly complex task,  requiring both a deep understanding of the design functionality and specialized expertise in SVA.
    \item Currently, there is a lack of a universal evaluation method and benchmarks capable of addressing the diverse types of VLSI designs.
\end{enumerate}

To tackle the identified challenges in SVA generation, in our work, we propose AssertLLM, a novel automatic assertion generation framework incorporating multiple specialized LLMs to deal with the decomposed tasks separately.
This framework is designed to process complete natural language specification files, automatically producing SVAs for each architectural signal. This approach significantly benefits both design-time bug prevention and verification-time bug detection. The role of AssertLLM within the standard VLSI design and verification flow is illustrated in Fig.~\ref{fig:motivation}. 
AssertLLM effectively addresses the outlined challenges by combining three customized LLMs, each focused on a specific task: extracting relevant information from specifications, mapping signal definitions, and translating natural language specifications into SVAs. The resulting SVAs contain various types, including bit-width, connectivity, and functional assertions.

Additionally, our work provides an open-source benchmark, designed to evaluate the quality of the generated SVAs. This benchmark, coupled with a general evaluation method, is adaptable to various design types.

To the best of our knowledge, AssertLLM is the first automatic assertion generation method that can handle full-size specification files and generate various types of SVAs for each architectural signal. It also provides the first open-source benchmark for assertion generation and evaluation
from design specifications, which can deal with different design types.

Our contributions in this work are summarized below:
\begin{itemize}
    \item To the best of our knowledge, AssertLLM is the first automatic assertion generation method that can handle the complete specification files and generate comprehensive types of SVAs for each architectural signal.
    \item We incorporate three customized LLMs, each enhanced with specific techniques for the decomposed tasks: extracting structural information from specifications, mapping signal declarations, and translating specifications into various SVA types. These SVAs support checks for bit-width, connectivity, and function.
    \item We provide the first open-source benchmarks\footnote{It will be open-sourced in https://github.com/hkust-zhiyao/AssertLLM} for assertion generation and evaluation, which include both golden specification documents and golden RTL designs. The generated SVAs are evaluated on the golden RTL implementations using model checking tools. Our evaluation method is designed to be applicable across a variety of design types.
    \item To demonstrate the effectiveness of AssertLLM, we conducted a comprehensive evaluation on a complete design. This resulted in the generation of 56 SVAs for 23 signals, with 23 for bit-width, 16 for interface, and 17 for function. Impressively, 89\% of these generated SVAs are evaluated to be correct both syntactically and functionally.
    
\end{itemize}

%% file: _txt/2_preliminaries.tex
\section{Preliminaries and Problem Formulation}

\subsection{Natural Language Specification}
\label{sec:nlspec}
A well-defined natural language specification mainly contains the following six parts:
(1) introduction: introduces the concepts and the features of the target design. (2) IO ports: provides detailed information on the prime input and prime output ports essential for interfacing. (3) registers: describe all the architecture-level registers in the design. (4) operation: explains the operational procedures for dataflow and control. (5) architecture: the high-level workflow and dataflow of the design. (6) usage examples: offers basic usage scenarios and corresponding waveform illustrations for the design. 

Specifically for the signals, the specification only defines the necessary architecture-level IO ports and registers, while leaving the definition of the internal signals used in the detailed RTL implementations for the RTL designers.

\subsection{LLM for EDA}
Recent advancements in LLMs like ChatGPT~\cite{achiam2023gpt} have not only demonstrated remarkable capability in content generation but also evolved to assist humans in various roles as agents. The application of LLMs in the field of electronic design automation (EDA) is an emerging area of exploration. Besides employing LLMs for assertion generation~\cite{kande2023llm, orenes2023using, sun2023towards}, recent studies have employed LLMs for tasks such as RTL code generation~\cite{liu2023rtlcoder, blocklove2023chip, lu2023rtllm, liu2023verilogeval, thakur2023benchmarking, thakur2023autochip, nair2023generating, liu2023chipnemo} and syntax correction~\cite{tsai2023rtlfixer}. Additionally, LLM-based solutions have been developed to facilitate the interaction with EDA tools~\cite{liu2023chipnemo, he2023chateda}, design architecture for AI accelerators~\cite{fu2023gpt4aigchip, yan2023viability}, fix security bugs~\cite{ahmad2023fixing}, generate and review specification documents~\cite{li2024specllm}, etc. These diverse applications and research efforts indicate a highly promising future for LLMs in enhancing and revolutionizing chip design processes.

\subsection{Problem Fromulation}

We denote a well-defined specification file as $\mathcal{S}$, where each architectural signal detailed in the specification is denoted as $sg_i$. Note that the architectural signal contains both the input and output ports and the architecture-level registers, but excludes the internal signals further implemented in the RTL design. Our generation process, denoted as $Gen$, is designed to analyze the specification file $\mathcal{S}$ and generate a comprehensive set of assertions $\mathcal{A}$ for each signal $sg_i$. The assertion generation can be expressed as follows:

\textbf{Problem 1} (Assertion Generation from Specification). 
\begin{equation}
    {\forall} sg_i \in \mathcal{S}, Gen(S, sg_i) \rightarrow \ \mathcal{A}(sg_i)
\end{equation}

Following the generation of assertions, it is crucial to evaluate their quality. We denote this evaluation process as $Eval$. To assess the correctness of the generated assertions, we utilize the golden RTL implementations, symbolized as $\mathcal{R}$. The correctness of the assertions is denoted as $Correct$. The evaluation can be formulated below:

\textbf{Problem 2} (Generated Assertion Evaluation). 
\begin{equation}
    {\forall} sg_i \in \mathcal{S}, Eval(\mathcal{R}, \mathcal{A}(sg_i)) \rightarrow \ Correct(\mathcal{A}(sg_i))
\end{equation}

%% file: _txt/3_methodology.tex
\section{Methodology}

\subsection{Workflow Overview}

\begin{figure*}[!ht]
  \centering
  \includegraphics[width=1\linewidth]{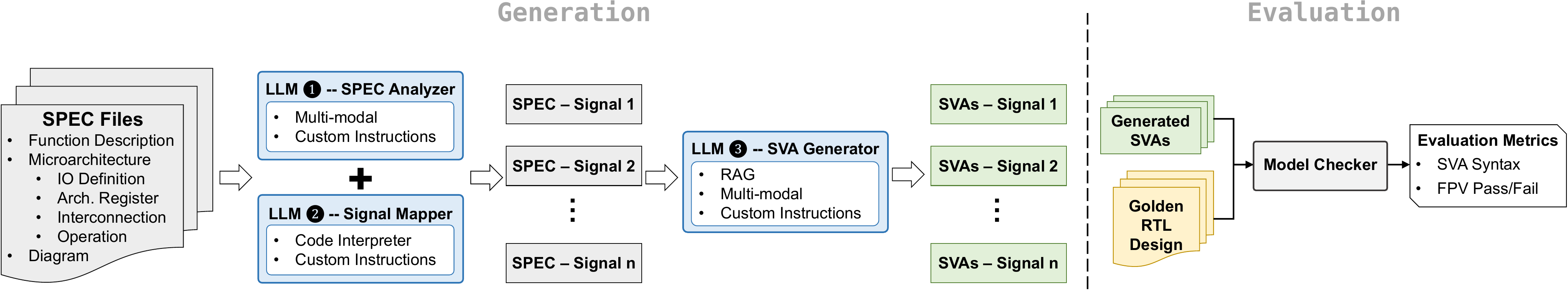}
      \caption{AssertLLM generation and evaluation workflow.  AssertLLM incorporates three customized LLMs, each enhanced with specific techniques for the decomposed tasks: extracting structural information from specifications, mapping signal definitions, and translating specifications into various SVA types. To evaluate the performance of the generation methods, the generated SVAs are further assessed based on the golden RTL implementations using model checking tools.}
  \label{fig:workflow}
  \vspace{-.2in}
\end{figure*}

Fig.~\ref{fig:workflow} illustrates the SVA generation and evaluation flow for AssertLLM.
Our approach to generating hardware verification assertions from natural language specifications, particularly from comprehensive specification documents, involves the integration of three customized LLMs. These LLMs are designed to break down this complex task into manageable components, thereby facilitating a comprehensive generation workflow. Additionally, we contribute an open-source benchmark and establish evaluation methodologies to assess the quality of the generated SVAs.

The assertion generation process is decomposed into three primary steps: (1) Extraction of relevant information from the original specification necessary for SVA generation. (2) Alignment of signal names between the natural language specifications and their corresponding declarations in HDL code. (3) Generation of high-quality SVAs based on the extracted natural language specifications.

In the subsequent subsections, we will detail the functionalities of each customized LLM of the comprehensive assertion generation flow. Following this, our SVA evaluation methodology will be presented.

\subsection{Specification Information Extraction}
The first step of our AssertLLM framework is to extract structured information from natural language specification documents to enable SVA generation. 
As we discussed in Section~\ref{sec:intro}, the first key challenge of SVA generation lies in the inherent unstructured nature of the original specifications, which contain background information, functional descriptions, microarchitecture designs, and various diagrams, including dataflow and waveform, etc. Meanwhile, the existence of assertion-relevant information across different sections further complicates the direct utilization of the original specifications for SVA generation.

Facing this challenge, existing methods can only deal with sentence-level specifications. Some works~\cite{frederiksen2020automated, parthasarathy2021spectosva} utilize ML methods to assess the relevance of manually extracted specification sentences to the intended assertions.  Other studies~\cite{harris2016glast, krishnamurthy2019controlled, zhao2019automatic, keszocze2019chatbot, aditi2022hybrid} directly use the human-identified or human-written assertion-related sentences. This reliance on sentence-level analysis limits the ability of the above NLP-based methods to fully automate SVA generation for practical applications

To address the challenge of processing original, unstructured, full-size specification documents, we propose a customized LLM, tailored to extract structural and relevant information for each defined signal, thereby further facilitating the SVA generation process.

\begin{figure}[!h]
  \centering
  \includegraphics[width=1\linewidth]{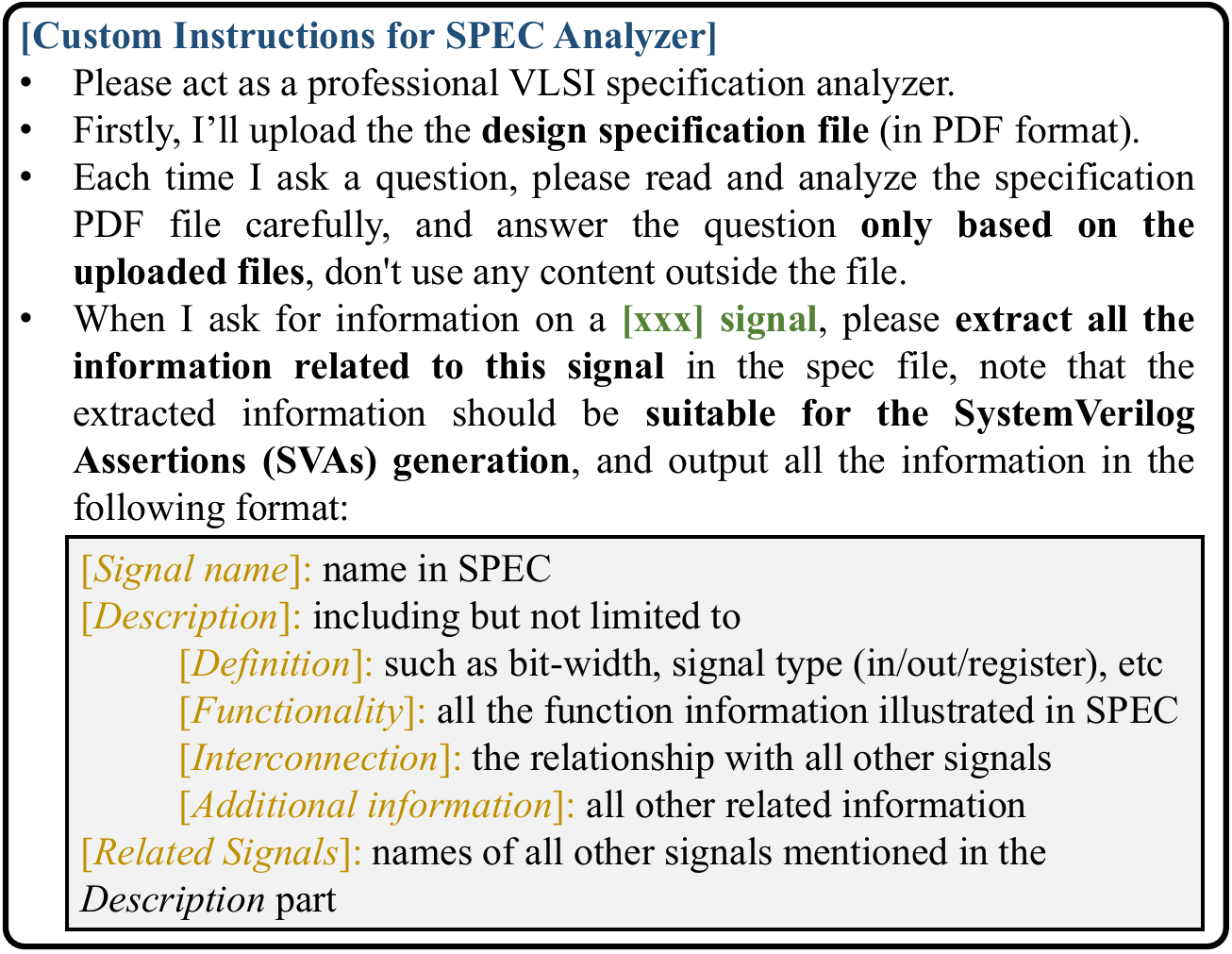}
  \caption{Custom Instructions for LLM \protect\circled{1}\protect \texttt{SPEC Analyzer}}
  \label{fig:llm1_inst}
  \vspace{-.2in}
\end{figure}

Specifically, in our LLM \circled{1} \texttt{SPEC Analyzer}, we first utilize system instructions to customize the LLM, shown as Fig.~\ref{fig:llm1_inst}.
The model takes the full-size specification file as the input, and the multi-modal function is employed to analyze the file containing text, table, figures, etc. 
Then for each signal, the LLM is required to extract all the related information of the signal. Here, we design a structured template to guide the LLM in extracting all essential signal-related information. This template contains three key components: the signal's name, its description, and the interconnection signals. We demonstrate the details of each part as follows:
\begin{itemize}
    \item \textbf{Name}: The identifier of the signal in the specification, ensuring clear and unambiguous reference.
    \item \textbf{Description}: To facilitate SVA generation, we divide the descriptions into four categories, including (1) definitions such as bit-width and signal type. (2) functionality which contains all the function-related information of the target signal in the entire specification file. (3) interconnection relationship with all other signals. (4) additional information that is not included in the above three types. 
    \item \textbf{Interconnection Signals}: A list of signals that interact or are associated with the target signal, which are essential for the assertion generation, and will be processed in the next LLM.
\end{itemize}
Note that the extracted information is summarized across different sections of the original specification, which contains all the information needed for assertion generation.

\begin{figure}[!h]
  \centering
  \includegraphics[width=1\linewidth]{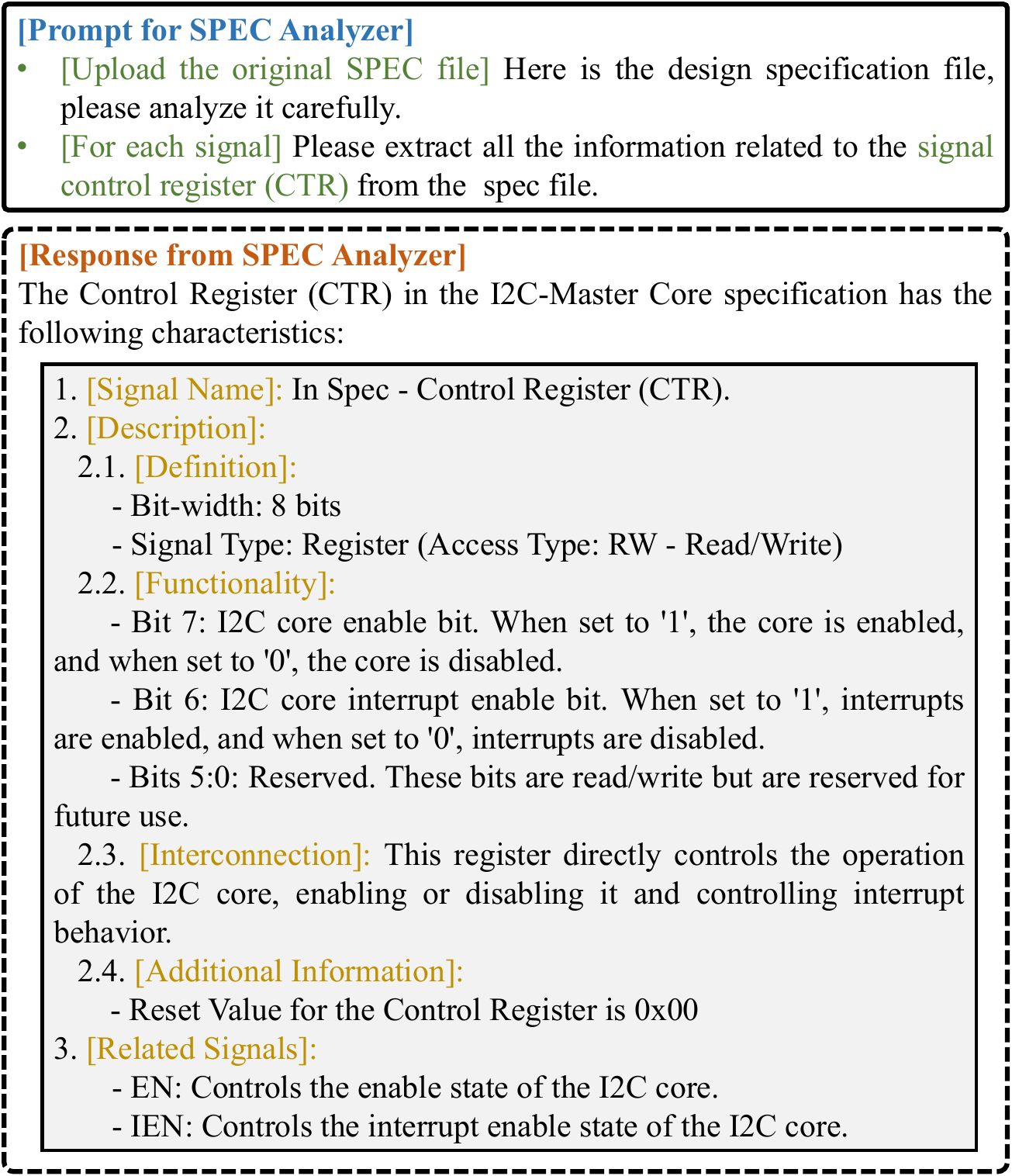}
  \caption{Prompt and Response Example of LLM \protect\circled{1}\protect \texttt{SPEC Analyzer}}
  \label{fig:llm1_prompt}
  \vspace{-.2in}
\end{figure}

\subsection{Signal Definition Mapping}

After extracting the structural specification information, we also face another problem: the target assertion contains internal signals that are not clearly defined in the specification file. As we illustrated in Subsection~\ref{sec:nlspec}, the specification document typically details only the input/output ports and architecture-level registers, while leaving the internal signals used to implement the detailed functions in RTL code (e.g., internal wires and registers) undefined.

To solve this problem, we introduce the second customized LLM \circled{2} \texttt{Signal Mapper} to analyze the signal definitions in the initialized HDL code and align these signal declarations with the natural language signal names found in the specification document.

\begin{figure}[!h]
  \centering
  \includegraphics[width=1\linewidth]{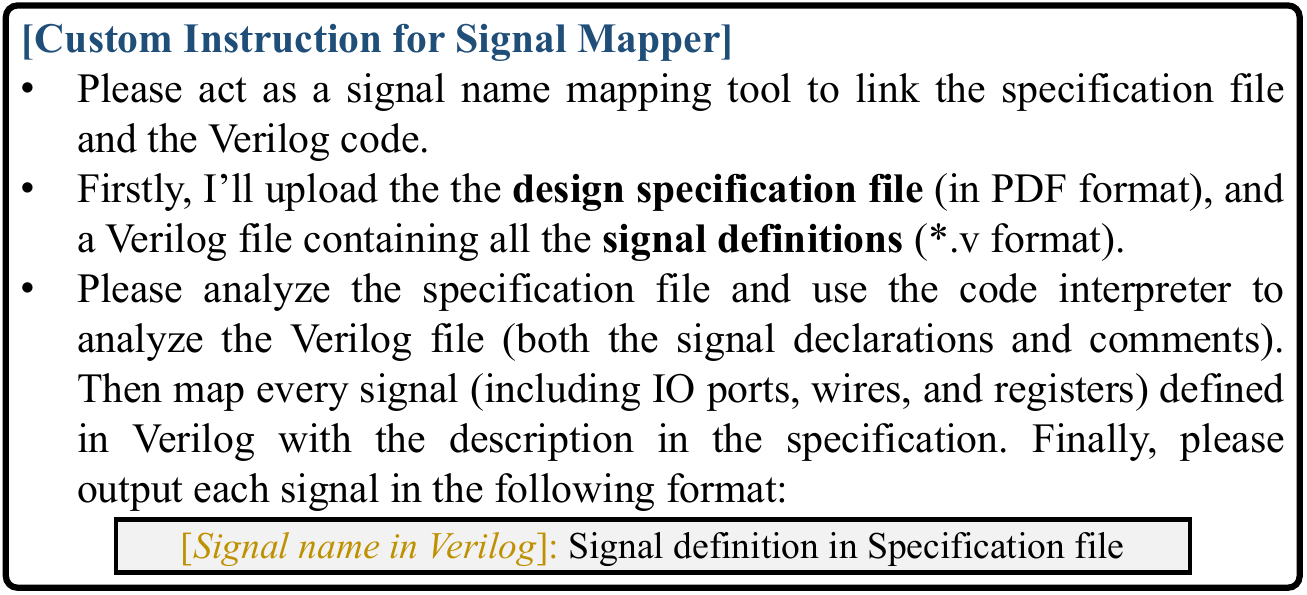}
  \caption{Custom Instructions for LLM \protect\circled{2}\protect \texttt{Signal Mapper}}
  \label{fig:llm2_inst}
  \vspace{-.2in}
\end{figure}

Specifically, we also use the custom instructions to guide LLM to specialize in the mapping task, shown in Fig.~\ref{fig:llm2_inst}. 
The model processes both the original specification file and the signal definition HDL code as inputs. It employs a code interpreter to carefully examine both the declarations and the comments within the HDL code snippet. Subsequently, the LLM analyzes the contents of the two files to establish a mapping relationship between the specification and the HDL code.

\begin{figure}[!h]
  \centering
  \includegraphics[width=1\linewidth]{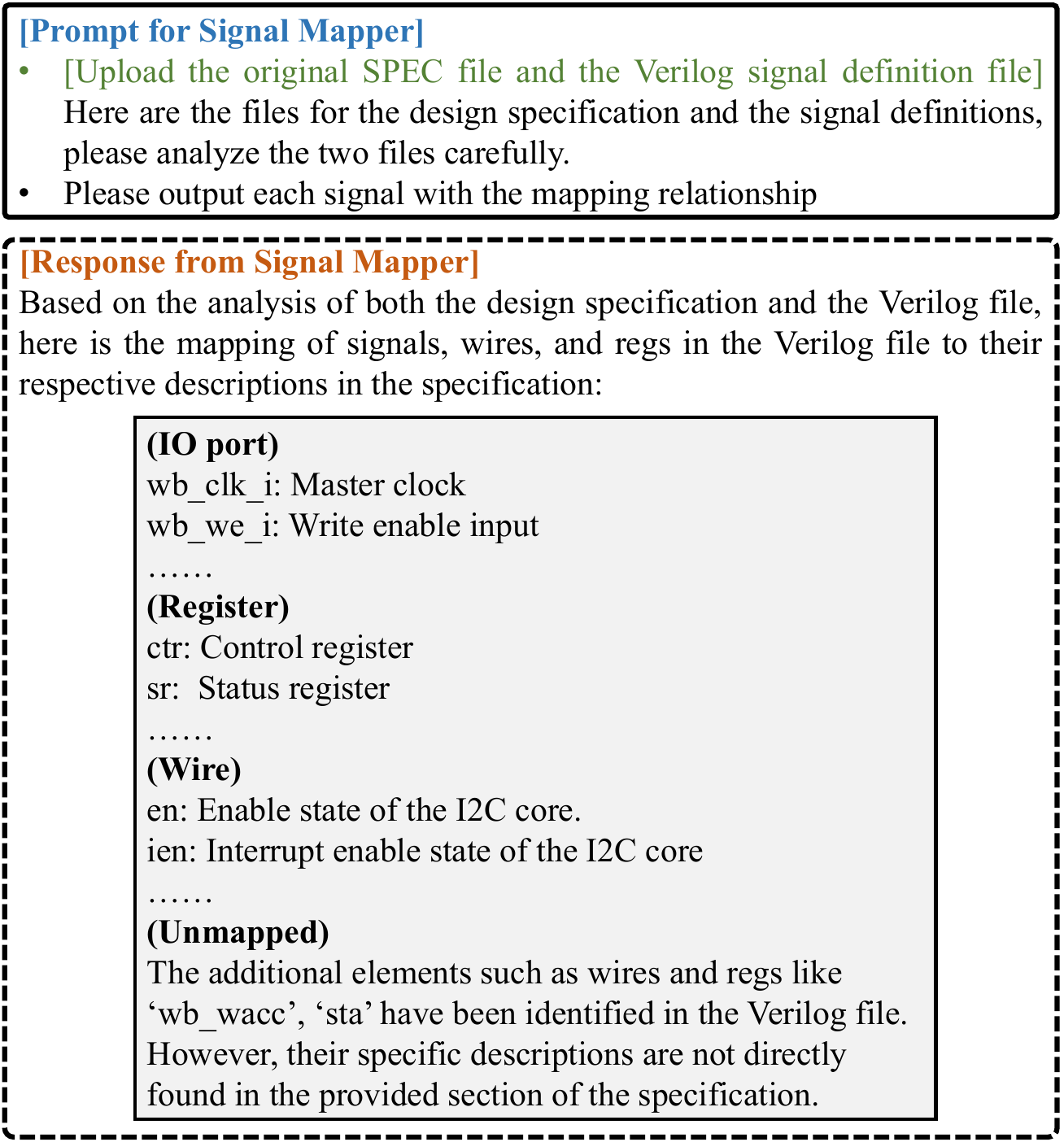}
  \caption{Prompt and Response Example of LLM \protect\circled{2}\protect \texttt{Signal Mapper}}
  \label{fig:llm2_prompt}
  \vspace{-.2in}
\end{figure}

\subsection{Automatic Assertion Generation}


While previous research has delved into SVA generation using either traditional NLP techniques at the pre-RTL stage or LLM-based approaches for RTL designs, these methods have their limitations. NLP-based techniques demand careful analysis of the syntax and semantics of assertion-related sentences, which limit their adaptability to variations in sentence structures. LLM-based methods, typically focused on the RTL stage, rely on HDL code and accompanying human-written comments or properties, but their dependence on the unverified RTL code poses a risk of generating inaccurate SVAs that could mislead the verification process.

To address these challenges, our work introduces the LLM \circled{3} \texttt{SVA Generator}, dedicated to generating assertions for each signal utilizing the previously extracted structural specifications and the established signal relationships.

Considering the precise syntax and writing rules inherent to SVAs and the potential for the original LLM failing to generate syntactically correct SVAs, as discussed in~\cite{orenes2023using}, we incorporate the Retrieval Augmented Generation (RAG) technique to enhance the LLM's capability for SVA generation. This approach is enriched by a knowledge database comprising tutorials and textbooks on SVA and formal property verification~\cite{seligman2023formal, mehta2020systemverilog, vijayaraghavan2005practical}, providing a robust foundation for the LLM to access and retrieve relevant SVA knowledge based on the input query, thereby enhancing the quality of the generated SVAs.

\begin{figure}[!h]
  \centering
  \includegraphics[width=1\linewidth]{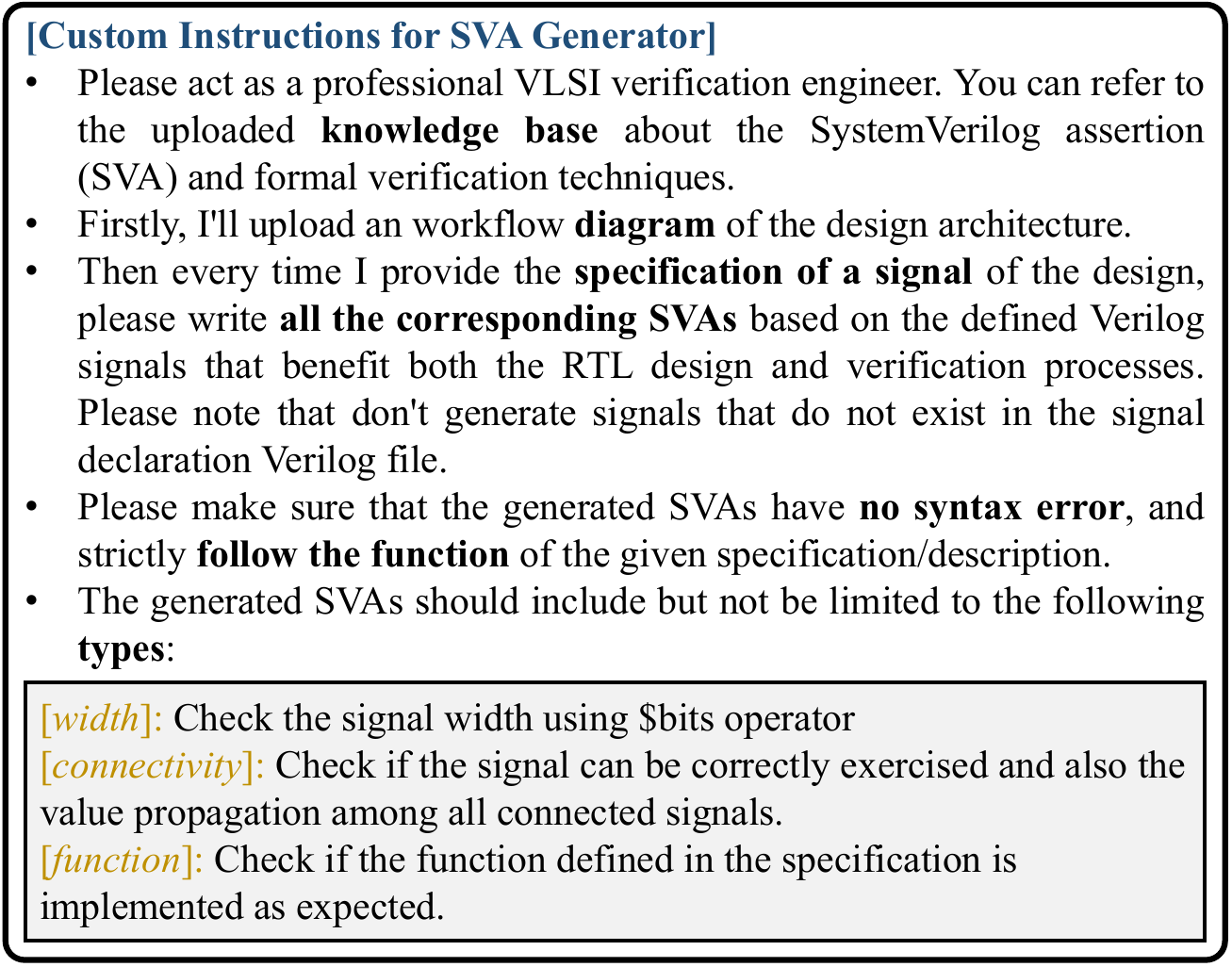}
  \caption{Custom Instructions for LLM \protect\circled{3}\protect \texttt{SVA Generator}}
  \label{fig:llm3_inst}
  \vspace{-.1in}
\end{figure}

Besides the RAG technique, we also provide custom instructions for \texttt{SVA Generator}, shown in Fig.~\ref{fig:llm3_inst}. After uploading the overall architecture diagram of the design, for each signal, the extracted structural specifications and the mapped signal relationship from the above two LLMs are provided. Then the LLM is required to generate SVAs strictly according to the specification, and as much and high coverage as possible. To guide the LLM to generate high-quality SVAs, we define five SVA categories as follows:

In addition to the RAG technique, we improve the \texttt{SVA Generator} with custom instructions, as illustrated in Fig.~\ref{fig:llm3_inst}. Upon inputting the overall architecture diagram of the design, the LLM is provided with the structured specifications and mapped signal relationships from the previous LLMs for each signal. Then the LLM is required to generate SVAs that adhere strictly to the specifications, aiming for maximal quantity and quality. To facilitate the generation of high-quality SVAs, we categorize SVAs into three distinct groups, guiding the LLM toward producing comprehensive and accurate assertions for effective verification.

\begin{figure}[!h]
  \centering
  \includegraphics[width=1\linewidth]{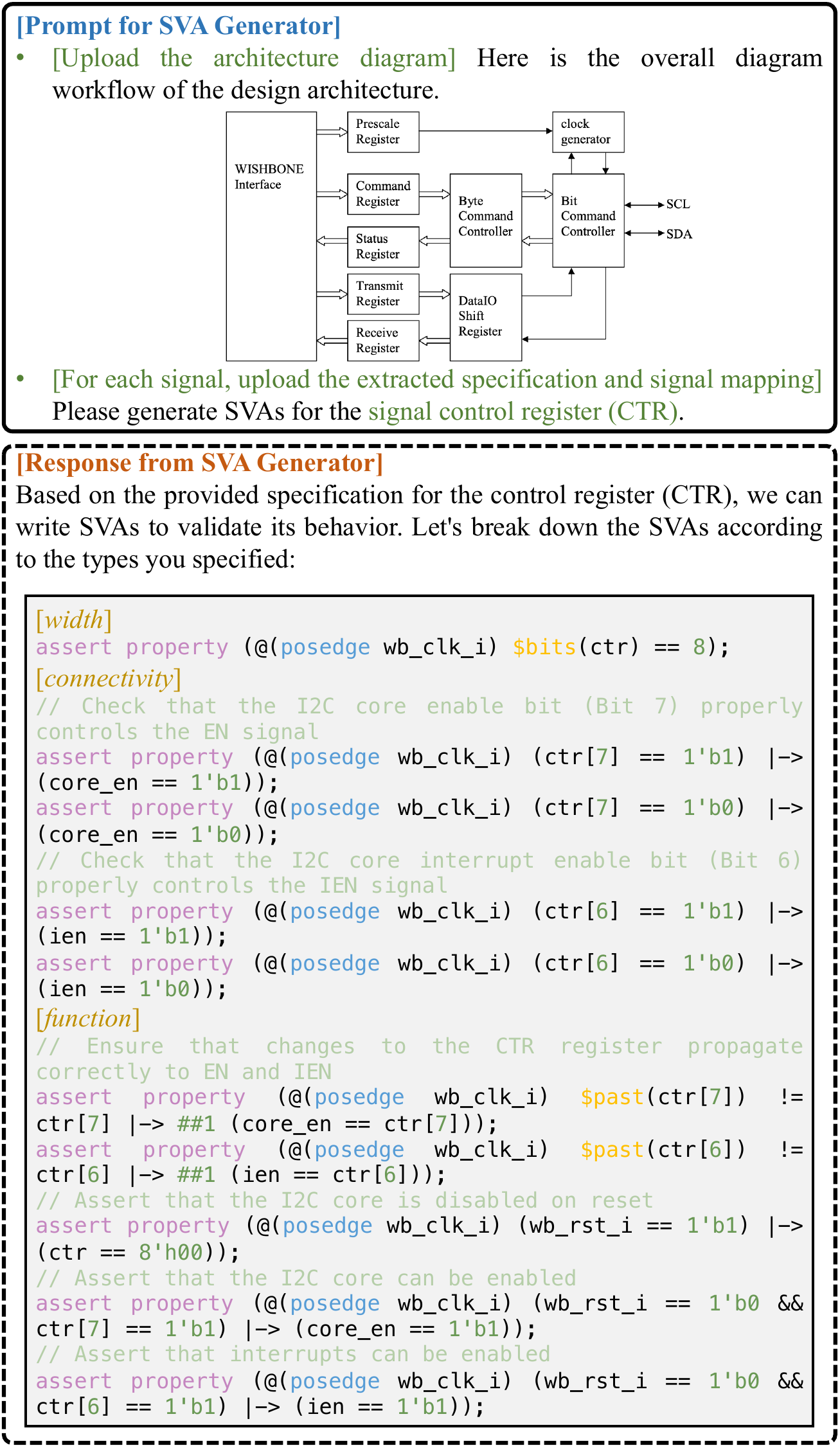}
  \caption{Prompt and Response Example of LLM \protect\circled{3}\protect \texttt{SVA Generator}}
  \label{fig:llm3_prompt}
  \vspace{-.2in}
\end{figure}

\begin{itemize}
    \item \textbf{Width}: Check if the signal bit width is satisfied with the specification.
    \item \textbf{Connectivity}: Check if the signal can be correctly exercised and also the value propagation among all connected signals.
    \item \textbf{Function}: Check if the function defined in the specification is implemented as expected.
\end{itemize}
Based on these well-designed SVA types, the customized LLM can generate numerous SVAs for each signal. Fig.~\ref{fig:llm3_prompt} demonstrates an example of generating SVAs for a signal.

\subsection{Generated Assertion Evaluation}

After the SVAs are generated, evaluating their effectiveness is crucial. While some previous studies such as~\cite{zhao2019automatic, frederiksen2020automated} suggest using specific property checkers for this purpose, such an approach is limited to particular design types like protocols and processors and lacks generalizability to all VLSI designs. Other methods like~\cite{orenes2023using} involve manual verification by engineers using EDA tools, which is vulnerable to human error.

In our approach, we propose leveraging the golden RTL implementations to assess the quality of the generated SVAs. Our generation method is solely based on the specification file, and the bug-free golden RTL designs serve as a robust benchmark to evaluate our generation technique's efficacy.

For evaluation, we utilize the formal property verification (FPV) method. The generated SVAs and the golden RTL designs are inputted into a model checker tool. After executing FPV, we employ the following two metrics to evaluate the quality of SVAs for each target signal:

\begin{itemize}
    \item \textbf{Syntax}: Check if the generated SVAs have syntax errors.
    \item \textbf{FPV Pass/Fail}: Given the RTL designs are bug-free, an SVA that passes the FPV check is considered correct, and conversely, a failure indicates an incorrect SVA.
\end{itemize}

\subsection{Proposed Benchmark}

Recognizing the absence of open-source benchmarks for evaluating LLMs in the generation of SVAs from natural language specifications, we introduce a comprehensive benchmark suite tailored for this purpose. Our benchmark consists of 20 open-source designs, covering a diverse array of applications including microprocessors, system-on-chip architectures, communication protocols, arithmetic units, and cryptographic modules. For each design within the benchmark, the benchmark provides the following components across three distinct files:

\begin{itemize}
    \item \textbf{Specification}: This file contains the complete natural language specification for the design, offering a detailed description of the system's intended architecture and functionality.
    \item \textbf{Signal Definition}: Presented in HDL code format, this file outlines the signal declarations essential for the generation of SVAs. It includes definitions for both input/output ports and internal signals, providing the foundation for assertion generation.
    \item \textbf{Golden RTL Implementation}: This file comprises the RTL design implementations that are strictly implemented according to the specification. The designs are verified to ensure it is free from bugs, serving as a reliable standard for evaluating the accuracy and effectiveness of generated SVAs.
    
\end{itemize}

%% file: _txt/4_experiment.tex
\section{Experimental Results}

\subsection{Experimental Setup}

In our study, the original specification documents are provided in PDF format, including a variety of multi-modal content including text, tables, and figures. The signal definition files and the golden RTL designs are formatted in Verilog. To assess the quality of the generated SVAs, we utilize Cadence JasperGold\textsuperscript{\textregistered}, one of the leading commercial model checking tools. This evaluation leverages the FPV app in JasperGold to ensure a thorough analysis.

Our experimental setup involves the evaluation of three types of LLMs using our developed generation and evaluation methodology:
\begin{enumerate}
    \item GPT-3.5: This model is the freely available commercial version, GPT-3.5 Turbo, which supports a context window of up to 16K tokens.
    \item GPT-4: The state-of-the-art commercial solution, GPT-4 Turbo, offers a 128K token context window and multi-modal capabilities, making it adept at handling the diverse content found in specification documents.
    \item AssertLLM:   Cutomized GPT-4 Turbo by incorporating specialized techniques such as RAG and custom instructions, tailoring the models specialized to the SVA generation task.
\end{enumerate}

In our experimental evaluation, we focus on the quality of the SVAs generated for each signal across the designs. Note that all SVAs are produced from a single query to the LLMs without any subsequent iterative modifications. The SVA containing unmapped signals is considered an unsuccessful attempt at SVA generation. These SVAs are identified and excluded by human engineers prior to the evaluation process.

\begin{table*}[!t]
\resizebox{0.995\textwidth}{!}{

\begin{tabular}{c|c|c||c|c|c|c||c||c}
\toprule
\multicolumn{3}{c||}{}                                                            & \multicolumn{4}{c||}{\textbf{AssertLLM}}                                & \textbf{GPT-4}             & \textbf{GPT-3.5}                                                                                                               \\ \hline
\multicolumn{3}{c||}{\textbf{Signal}}                                             & \multicolumn{6}{c}{\textbf{Assertion Evaluation (\#. Generated/\#. Syntax Correct/\#. FPV Pass)}}                                                                                                                                \\ \hline 
\multicolumn{2}{c}{\textbf{Type}}                           & \textbf{Name}     & \textbf{Width} & \textbf{Connect.}   & \textbf{Function}    & \textbf{Signal Total} & \textbf{Function} &                                                                                                                       \\ \hline \hline
                           & \textit{Clock}                          & wb\_clk\_i        & 1/1/1          & \multicolumn{2}{c|}{/}                       & 1/1/1                 & 3/1/0             &                                                                                                                       \\ \cline{2-8}
                           &                                & wb\_rst\_i        & 1/1/1          & \multicolumn{2}{c|}{}                        & 1/1/1                 & 3/1/0             &                                                                                                                       \\
                           & \multirow{-2}{*}{\textit{Reset}}        & arst\_i           & 1/1/1          & \multicolumn{2}{c|}{\multirow{-2}{*}{/}}     & 1/1/1                 & 3/1/0             &                                                                                                                       \\ \cline{2-8}
                           &                                & wb\_stb\_i        & 1/1/1          & 2/2/1                &                      & 3/3/2                 & 3/1/0             &                                                                                                                       \\
                           &                                & wb\_ack\_o        & 1/1/1          & 1/1/0                &                      & 2/2/1                 & 3/1/0             &                                                                                                                       \\
                           & \multirow{-3}{*}{\textit{Control}}      & wb\_inta\_o       & 1/1/1          & 1/1/0                &   \multirow{-3}{*}{/}                   & 2/2/1                 & 3/1/0             &                                                                                                                       \\ \cline{2-8}
                           &                                & wb\_adr\_i        & 1/1/1          &                      &                      & 1/1/1                 & 3/1/0             &                                                                                                                       \\
                           &                                & wb\_dat\_i        & 1/1/1          &                      &                      & 1/1/1                 & 3/1/0             &                                                                                                                       \\
                           &                                & wb\_cyc\_i        & 1/1/1          &                      &                      & 1/1/1                 & 3/1/0             &                                                                                                                       \\
                           &                                & wb\_dat\_o        & 1/1/1          &                      &                      & 1/1/1                 & 3/1/0             &                                                                                                                       \\
                           &                                & wb\_we\_i         & 1/1/1          &                      &                      & 1/1/1                 & 3/1/0             &                                                                                                                       \\
                           &                                & scl\_pad\_i       & 1/1/1          &                      &                      & 1/1/1                 & 3/1/0             &                                                                                                                       \\
                           &                                & scl\_pad\_o       & 1/1/1          &                      &                      & 1/1/1                 & 3/1/0             &                                                                                                                       \\
                           &                                & sda\_pad\_i       & 1/1/1          &                      &                      & 1/1/1                 & 3/1/0             &                                                                                                                       \\
                           &                                & sda\_pad\_o       & 1/1/1          &                      &                      & 1/1/1                 & 3/1/0             &                                                                                                                       \\
                           &                                & scl\_pad\_oe      & 1/1/1          &                      &                      & 1/1/1                 & 3/1/0             &                                                                                                                       \\
\multirow{-17}{*}{IO Port} & \multirow{-11}{*}{\textit{Data}}        & sda\_pad\_oe      & 1/1/1          & \multirow{-11}{*}{/} & \multirow{-11}{*}{/} & 1/1/1                 & 3/1/0             &                                                                                                                       \\ \cline{1-8}
                           &                                & ctr               & 1/1/1          & 4/4/4                & 5/5/5                & 10/10/10              & 3/1/1             &                                                                                                                       \\
                           & \multirow{-2}{*}{\textit{Control}}      & sr                & 1/1/1          & 6/6/5                & 8/8/8                & 15/15/14              & 3/1/1             &                                                                                                                       \\ \cline{2-8}
                           
                           &                                & prer              & 1/1/1          & /                    & 3/3/1                & 4/4/2                 & 4/1/1             &                                                                                                                       \\
                           &                                & txr               & 1/1/1          & /                    & 1/1/1                & 2/2/2                 & 3/1/1             &                                                                                                                       \\
                           &                                & rxr               & 1/1/1          & /                    & 1/1/1                & 2/2/2                 & 3/1/1             &                                                                                                                       \\
\multirow{-6}{*}{Register}  & \multirow{-4}{*}{\textit{Data}}         & cr                & 1/1/1          & /                    & 1/1/1                & 2/2/2                 & 4/1/1             &                                                                                                                       \\ \hline \hline
\multicolumn{1}{l}{}       & \multicolumn{2}{c||}{}                               & 23/23/23       & 16/16/12             & 17/17/15             & 56/56/50              & 71/23/6           &                                                                                                                       \\ 
\multicolumn{1}{l}{}       & \multicolumn{2}{c||}{\multirow{-2}{*}{\textbf{Design Total}}} & 100\%/100\%    & 100\%/75\%           & 100\%/88\%           & 100\%/89\%            & 32\%/8\%          & \multirow{-26}{*}{\begin{tabular}[c]{@{}c@{}}Can not handle \\ the  original \\ specification files.\end{tabular}}

\\ \bottomrule
\end{tabular}

}

\caption{Evaluation of the generated SVAs for design "I2C".  AssertLLM generates 56 properties for a total of 23 signals, with 23 for bit-width, 16 for connectivity, and 17 for function. 89\% of these generated SVAs are evaluated to be correct both syntactically and functionally.}
\label{tbl:eval}
 
\end{table*}

\subsection{Evaluation Metrics}

To conduct a thorough evaluation of the generated SystemVerilog SVAs, we propose a set of metrics that align with our evaluation methodology. This approach ensures a detailed assessment of the SVAs' quality on both a per-signal and per-design basis.

For each assertion type of an individual signal, our evaluation includes the following metrics: (1) number of generated SVAs. (2) number of syntax-correct SVAs. (3) number of FPV-passed SVAs. 

Once the evaluation for each signal is complete, we aggregate the statistics of the generated SVAs for each design and then calculate the proportion of these SVAs that are syntactically correct and passed the FPV checks, respectively.


\subsection{Assertion Generation Quality}

To illustrate the efficacy of AssertLLM, we apply it to a comprehensive design case: the "I2C" protocol. The I2C specification describes the architecture of a serial communication bus that provides a simple and efficient method of data exchange between devices. The complete specification document for the "I2C" design is structured into six main sections, similar to those illustrated in Subsection~\ref{sec:nlspec}.
Note that for each signal, the specification is unstructured and mainly across the IO ports, registers, and operation sections.

Additionally, we provide the signal definition file containing not only the IO ports and architectural registers but also all the internal wires and registers defined for detailed RTL implementation.

To facilitate the generation of SVAs, the AssertLLM framework processes specification and signal definition files using two specialized LLMs: SPEC Analyzer for extracting structured specifications for each signal and Signal Mapper for mapping signal relationships. Then the SVA Generator is utilized to automatically generate three types of SVAs based on the processed information from the first two LLMs.

The specification for the "I2C" design defines 23 signals, comprising 17 IO ports and 6 architecture-level registers. For the IO ports, we categorize them into 4 functional types: clock, reset, control signal, and data signal. The architecture-level registers are similarly categorized, based on their functionality, into control and data types.

The evaluation of SVAs generated by our AsserLLM is demonstrated in Table~\ref{tbl:eval}. For each signal, we first verify each type of the generated SVAs separately. Then we summarize all the SVAs to provide a design-level inspection. We have multiple interesting observations in Table~\ref{tbl:eval} as follows:
\begin{itemize}
    \item AssertLLM demonstrates excellent proficiency in generating SVAs for bit-width checking. Although bit-width checking is relatively straightforward, it is crucial for early design stages to avoid potential bugs that cannot be checked through a syntax checker.
    \item For the connectivity SVAs, since clear guidelines are provided only for control signals within the architecture-level specification documents, AssertLLM can only generate connectivity SVAs for them. The connectivity of data signals often depend on specific internal signals defined in the RTL implementation, which are not detailed in the specification document.
    \item For the function SVAs, the specification provides explicit details only for registers. The descriptions of IO ports  primarily focus on data transformation functions, without extensive functional details, which results in the lack of related SVAs.
    \item For the quantity of generated SVAs, AssertLLM produced a total of 56 SVAs, with 23 dedicated to width checking, 16 to connectivity checking, and 17 to function checking. 
    \item Regarding the quality of generated SVAs, all SVAs related to bit-width checking performed correctly. However, a minor portion of connectivity and function SVAs contained errors, attributed mainly to misinterpretations of the specification or LLM-generated hallucinations. Overall, AssertLLM achieved a correct accuracy rate of 89\% for the entire I2C design.
    
\end{itemize}

\subsection{Ablation Study}

In addition to assessing AssertLLM's performance, we conducted an ablation study to compare the SVA generation capabilities of the original GPT-4 and GPT-3.5 models without the additional techniques. This study provides insights into the effectiveness of enhancements incorporated in AssertLLM.

The evaluation results for the two commercial solutions are demonstrated in Table~\ref{tbl:eval}. For GPT-3.5, since the lack of multi-modal processing capabilities, it is unable to directly generate SVAs from the original, multi-modal specification files. 

When utilizing the original GPT-4, the unstructured specification file and signal definitions are provided, with prompts designed to guide SVA generation to the best of the model's ability. The generation results indicate that the absence of a mechanism to extract structured specifications for each signal significantly hampers GPT-4's ability to compile all useful information for SVA generation, resulting in a maximum of only 4 SVAs generated per signal.
Additionally, without specific assertion type guidance, GPT-4 only generates functional assertions.

For the SVA quality of commercial solutions, the original GPT-4 model tends to produce SVAs with syntax errors, similar to observations made in the previous study~\cite{orenes2023using}. This issue is addressed in AssertLLM through the application of RAG techniques, which enrich the model with specific knowledge on SVA and Formal Property Verification (FPV). In the evaluation results, GPT-4 failed to generate any correct SVAs for IO ports and only succeeded in creating accurate reset check assertions for registers, leading to an overall correct proportion of only 8\%.

%% file: _txt/5_discussion.tex
\section{Discussion}

\subsection{Coverage in SVA Evaluation}
Some works~\cite{orenes2023using} propose to leverage the coverage metric, especially the cone-of-influence (COI) coverage to evaluate the quality of generated SVAs. 
COI coverage relies on analyzing the signals exercised during simulation or formal verification, which significantly involves internal signals within the design.

Given that our SVA generation process is based solely on the information available in the specification documents, which typically detail external interfaces like IO ports and architectural-level registers rather than internal signals, COI coverage does not align well with our evaluation criteria. This coverage metric assumes a level of design implementation detail that goes beyond the scope of natural language specifications, making it less applicable for assessing the completeness or effectiveness of SVAs generated at this pre-RTL stage.

\subsection{Evaluating and Enhancing Specification Quality with AssertLLM}

The generation of high-quality SVAs from natural language specifications relies not only on the capabilities of LLMs but also on the intrinsic quality of the specification documents themselves. A specification that provides only the basic information of signals, such as their names and simple descriptions, without delving into detailed functionalities or connectivities, inherently limits the potential for generating meaningful SVAs, regardless of the power of the LLMs employed. Conversely, specifications that offer comprehensive details, including clear definitions of signal functionalities and connectivities, can facilitate the generation of SVAs even with relatively simple LLMs.

Here we identify a novel application for AssertLLM beyond its primary role in verification: utilizing AssertLLM as a tool for assessing the quality of natural language specifications. This application leverages AssertLLM's ability to process and interpret specification documents to determine their \textit{verification-friendliness}. Specifications that enable AssertLLM to generate a broad and accurate range of SVAs can be considered high-quality and well-suited for verification purposes. This approach to evaluating specification quality offers several benefits:

\begin{itemize}
    \item \textbf{Identifying Gaps between Specifications and Verification}: AssertLLM can highlight contents within a specification that lack sufficient detail for SVA generation, guiding architects to provide more comprehensive information.
    \item \textbf{Enhancing Verification}: Ensuring specifications are verification-friendly can potentially reduce the time and effort required for verification.
    \item \textbf{Standardizing Specification Writing}: The feedback from AssertLLM can help establish best practices for writing specifications to facilitate the automated verification, and promote consistency across design stages.
    
\end{itemize}

%% file: _txt/6_conclusion.tex
\section{Conclusion}

In this study, we introduce AssertLLM, an automated framework designed for generating assertions from entire specification documents. AssertLLM breaks down the intricate task into three sequential phases, leveraging specialized LLMs for structural specification extraction, signal definition mapping, and assertion creation. We also offer an open-source benchmark to evaluate the efficacy of assertion generation. Evaluating AssertLLM on a comprehensive design with 23 signals revealed that 89\% of the assertions generated were accurate both syntactically and functionally. We also discuss the potential of using AssertLLM to evaluate and enhance the quality of specifications.

%% file: assertion.bib
@inproceedings{harris2016glast,
  title={Glast: Learning formal grammars to translate natural language specifications into hardware assertions},
  author={Harris, Christopher B and Harris, Ian G},
  booktitle={2016 Design, Automation \& Test in Europe Conference \& Exhibition (DATE)},
  pages={966--971},
  year={2016},
  organization={IEEE}
}

@inproceedings{krishnamurthy2019controlled,
  title={Controlled natural language framework for generating assertions from hardware specifications},
  author={Krishnamurthy, Rahul and Hsiao, Michael S},
  booktitle={2019 IEEE 13th International Conference on Semantic Computing (ICSC)},
  pages={367--370},
  year={2019},
  organization={IEEE}
}

@inproceedings{zhao2019automatic,
  title={Automatic assertion generation from natural language specifications using subtree analysis},
  author={Zhao, Junchen and Harris, Ian G},
  booktitle={2019 Design, Automation \& Test in Europe Conference \& Exhibition (DATE)},
  pages={598--601},
  year={2019},
  organization={IEEE}
}

@inproceedings{krishnamurthy2019ease,
  title={Ease: Enabling hardware assertion synthesis from english},
  author={Krishnamurthy, Rahul and Hsiao, Michael S},
  booktitle={Rules and Reasoning: Third International Joint Conference, RuleML+ RR 2019, Bolzano, Italy, September 16--19, 2019, Proceedings 3},
  pages={82--96},
  year={2019},
  organization={Springer}
}

@inproceedings{frederiksen2020automated,
  title={Automated Assertion Generation from Natural Language Specifications},
  author={Frederiksen, Steven J and Aromando, John and Hsiao, Michael S},
  booktitle={2020 IEEE International Test Conference (ITC)},
  pages={1--5},
  year={2020},
  organization={IEEE}
}

@inproceedings{keszocze2019chatbot,
  title={Chatbot-based assertion generation from natural language specifications},
  author={Keszocze, Oliver and Harris, Ian G},
  booktitle={2019 Forum for Specification and Design Languages (FDL)},
  pages={1--6},
  year={2019},
  organization={IEEE}
}

@inproceedings{parthasarathy2021spectosva,
  title={SpecToSVA: Circuit Specification Document to SystemVerilog Assertion Translation},
  author={Parthasarathy, Ganapathy and Nanda, Saurav and Choudhary, Parivesh and Patil, Pawan},
  booktitle={2021 Second Document Intelligence Workshop at KDD}
}

@inproceedings{aditi2022hybrid,
  title={Hybrid Rule-based and Machine Learning System for Assertion Generation from Natural Language Specifications},
  author={Aditi, Fnu and Hsiao, Michael S},
  booktitle={2022 IEEE 31st Asian Test Symposium (ATS)},
  pages={126--131},
  year={2022},
  organization={IEEE}
}

@article{kande2023llm,
  title={LLM-assisted Generation of Hardware Assertions},
  author={Kande, Rahul and Pearce, Hammond and Tan, Benjamin and Dolan-Gavitt, Brendan and Thakur, Shailja and Karri, Ramesh and Rajendran, Jeyavijayan},
  journal={arXiv preprint arXiv:2306.14027},
  year={2023}
}

@article{orenes2023using,
  title={Using LLMs to Facilitate Formal Verification of RTL},
  author={Orenes-Vera, Marcelo and Martonosi, Margaret and Wentzlaff, David},
  journal={arXiv e-prints},
  pages={arXiv--2309},
  year={2023}
}

@inproceedings{sun2023towards,
  title={Towards Improving Verification Productivity with Circuit-Aware Translation of Natural Language to SystemVerilog Assertions},
  author={Sun, Chuyue and Hahn, Christopher and Trippel, Caroline},
  booktitle={First International Workshop on Deep Learning-aided Verification (DAV)},
  year={2023}
}

@book{seligman2023formal,
  title={Formal verification: an essential toolkit for modern VLSI design},
  author={Seligman, Erik and Schubert, Tom and Kumar, MV Achutha Kiran},
  year={2023},
  publisher={Elsevier}
}

@book{mehta2020systemverilog,
  title={SystemVerilog Assertions and Functional Coverage},
  author={Mehta, Ashok B},
  year={2020},
  publisher={Springer}
}

@book{vijayaraghavan2005practical,
  title={A practical guide for SystemVerilog assertions},
  author={Vijayaraghavan, Srikanth and Ramanathan, Meyyappan},
  year={2005},
  publisher={Springer Science \& Business Media}
}

@article{achiam2023gpt,
  title={GPT-4 Technical Report},
  author={Achiam, Josh and Adler, Steven and Agarwal, Sandhini and Ahmad, Lama and Akkaya, Ilge and Aleman, Florencia Leoni and Almeida, Diogo and Altenschmidt, Janko and Altman, Sam and Anadkat, Shyamal and others},
  journal={arXiv preprint arXiv:2303.08774},
  year={2023}
}

@article{lu2023rtllm,
  title={RTLLM: An open-source benchmark for design rtl generation with large language model},
  author={Lu, Yao and Liu, Shang and Zhang, Qijun and Xie, Zhiyao},
  journal={arXiv preprint arXiv:2308.05345},
  year={2023}
}

@article{blocklove2023chip,
  title={Chip-Chat: Challenges and Opportunities in Conversational Hardware Design},
  author={Blocklove, Jason and Garg, Siddharth and Karri, Ramesh and Pearce, Hammond},
  journal={arXiv preprint arXiv:2305.13243},
  year={2023}
}

@article{liu2023verilogeval,
  title={VerilogEval: Evaluating Large Language Models for Verilog Code Generation},
  author={Liu, Mingjie and Pinckney, Nathaniel and Khailany, Brucek and Ren, Haoxing},
  journal={arXiv preprint arXiv:2309.07544},
  year={2023}
}

@inproceedings{thakur2023benchmarking,
  title={Benchmarking Large Language Models for Automated Verilog RTL Code Generation},
  author={Thakur, Shailja and Ahmad, Baleegh and Fan, Zhenxing and Pearce, Hammond and Tan, Benjamin and Karri, Ramesh and Dolan-Gavitt, Brendan and Garg, Siddharth},
  booktitle={DATE},
  year={2023},
}

@article{thakur2023autochip,
  title={AutoChip: Automating HDL Generation Using LLM Feedback},
  author={Thakur, Shailja and Blocklove, Jason and Pearce, Hammond and Tan, Benjamin and Garg, Siddharth and Karri, Ramesh},
  journal={arXiv preprint arXiv:2311.04887},
  year={2023}
}

@article{nair2023generating,
  title={Generating secure hardware using chatgpt resistant to cwes},
  author={Nair, Madhav and Sadhukhan, Rajat and Mukhopadhyay, Debdeep},
  journal={Cryptology ePrint Archive},
  year={2023}
}

@article{liu2023rtlcoder,
  title={RTLCoder: Outperforming GPT-3.5 in Design RTL Generation with Our Open-Source Dataset and Lightweight Solution},
  author={Liu, Shang and Fang, Wenji and Lu, Yao and Zhang, Qijun and Zhang, Hongce and Xie, Zhiyao},
  journal={arXiv preprint arXiv:2312.08617},
  year={2023}
}

@article{liu2023chipnemo,
  title={ChipNeMo: Domain-Adapted LLMs for Chip Design},
  author={Liu, Mingjie and Ene, Teodor-Dumitru and Kirby, Robert and Cheng, Chris and Pinckney, Nathaniel and Liang, Rongjian and Alben, Jonah and Anand, Himyanshu and Banerjee, Sanmitra and Bayraktaroglu, Ismet and others},
  journal={arXiv preprint arXiv:2311.00176},
  year={2023}
}

@inproceedings{he2023chateda,
  title={ChatEDA: A Large Language Model Powered Autonomous Agent for EDA},
  author={He, Zhuolun and Wu, Haoyuan and Zhang, Xinyun and Yao, Xufeng and Zheng, Su and Zheng, Haisheng and Yu, Bei},
  booktitle={MLCAD Workshop},
  year={2023},
}

@article{ahmad2023fixing,
  title={Fixing Hardware Security Bugs with Large Language Models},
  author={Ahmad, Baleegh and Thakur, Shailja and Tan, Benjamin and Karri, Ramesh and Pearce, Hammond},
  journal={arXiv preprint arXiv:2302.01215},
  year={2023}
}

@article{fu2023gpt4aigchip,
  title={GPT4AIGChip: Towards Next-Generation AI Accelerator Design Automation via Large Language Models},
  author={Fu, Yonggan and Zhang, Yongan and Yu, Zhongzhi and Li, Sixu and Ye, Zhifan and Li, Chaojian and Wan, Cheng and Lin, Yingyan},
  journal={arXiv preprint arXiv:2309.10730},
  year={2023}
}

@article{yan2023viability,
  title={On the Viability of using LLMs for SW/HW Co-Design: An Example in Designing CiM DNN Accelerators},
  author={Yan, Zheyu and Qin, Yifan and Hu, Xiaobo Sharon and Shi, Yiyu},
  journal={arXiv preprint arXiv:2306.06923},
  year={2023}
}

@article{tsai2023rtlfixer,
  title={RTLFixer: Automatically Fixing RTL Syntax Errors with Large Language Models},
  author={Tsai, YunDa and Liu, Mingjie and Ren, Haoxing},
  journal={arXiv preprint arXiv:2311.16543},
  year={2023}
}

@article{li2024specllm,
  title={SpecLLM: Exploring Generation and Review of VLSI Design Specification with Large Language Model},
  author={Li, Mengming and Fang, Wenji and Zhang, Qijun and Xie, Zhiyao},
  journal={arXiv preprint arXiv:2401.13266},
  year={2024}
}

@article{germiniani2022harm,
  title={Harm: a hint-based assertion miner},
  author={Germiniani, Samuele and Pravadelli, Graziano},
  journal={IEEE Transactions on Computer-Aided Design of Integrated Circuits and Systems},
  volume={41},
  number={11},
  pages={4277--4288},
  year={2022},
  publisher={IEEE}
}

@inproceedings{danese2017team,
  title={A-team: Automatic template-based assertion miner},
  author={Danese, Alessandro and Riva, Nicol{\`o} Dalla and Pravadelli, Graziano},
  booktitle={Proceedings of the 54th Annual Design Automation Conference 2017},
  pages={1--6},
  year={2017}
}

@inproceedings{vasudevan2010goldmine,
  title={Goldmine: Automatic assertion generation using data mining and static analysis},
  author={Vasudevan, Shobha and Sheridan, David and Patel, Sanjay and Tcheng, David and Tuohy, Bill and Johnson, Daniel},
  booktitle={2010 Design, Automation \& Test in Europe Conference \& Exhibition (DATE 2010)},
  pages={626--629},
  year={2010},
  organization={IEEE}
}

@article{fang2023r,
  title={r-map: Relating Implementation and Specification in Hardware Refinement Checking},
  author={Fang, Wenji and Hu, Guangyu and Zhang, Hongce},
  journal={IEEE Transactions on Computer-Aided Design of Integrated Circuits and Systems},
  year={2023},
  publisher={IEEE}
}

@inproceedings{orenes2021autosva,
  title={AutoSVA: Democratizing Formal Verification of RTL Module Interactions},
  author={Orenes-Vera, Marcelo and Manocha, Aninda and Wentzlaff, David and Martonosi, Margaret},
  booktitle={2021 58th ACM/IEEE Design Automation Conference (DAC)},
  pages={535--540},
  year={2021},
  organization={IEEE}
}

@article{witharana2022survey,
  title={A survey on assertion-based hardware verification},
  author={Witharana, Hasini and Lyu, Yangdi and Charles, Subodha and Mishra, Prabhat},
  journal={ACM Computing Surveys (CSUR)},
  volume={54},
  number={11s},
  pages={1--33},
  year={2022},
  publisher={ACM New York, NY}
}
